\documentstyle[aps,psfig,prb,preprint]{revtex}

\begin{document}

 \title{Magnetic field effects on quantum ring excitons}

 \author{Jakyoung Song$^{a,b}$ and Sergio E. Ulloa$^{b,c}$}

 \address{$^a$National CRI Center for Nano
Particle Control, School of Mechanical and Aerospace Engineering,
Seoul
National University, Seoul, 151-742, Korea \\
$^b$Department of Physics and Astronomy and Condensed Matter and
  Surface Sciences Program, Ohio University, Athens, Ohio 45701--2979 \\
$^c$Sektion Physik, Ludwig-Maximilians-Universit\"at,
Geschwister-Scholl-Platz 1, 80539, Germany}

 \maketitle

\begin{abstract}

We study the effect of magnetic field and geometric confinement on
excitons confined to a quantum ring. We use analytical matrix
elements of the Coulomb interaction and diagonalize numerically
the effective-mass Hamiltonian of the problem.  To explore the
role of different boundary conditions, we investigate the quantum
ring structure with a parabolic confinement potential, which
allows the wavefunctions to be expressed in terms of center of
mass and relative degrees of freedom of the exciton. On the other
hand, wavefunctions expressed in terms of Bessel functions for
electron and hole are used for a hard-wall confinement potential.
The binding energy and electron-hole separation of the exciton
are calculated as function of the width of the ring and the
strength of a external magnetic field.  The linear optical
susceptibility as a function of magnetic fields is also
discussed.  We explore the Coulomb electron-hole correlation and
magnetic confinement for several ring width and size combinations.
The Aharanov-Bohm oscillations of exciton characteristics
predicted for one-dimensional rings are found to not be present
in these finite-width systems.
 \end{abstract}

 \pacs{71.35.Ji, 71.35.Cc, 71.35.Cc, 78.55.-m}
%\begin{multicols}{2}

 \narrowtext

 \section{ Introduction }

The fabrication of nanometer-size semiconductor structures by
different techniques (including lithography, etching, direct
chemical synthesis, and self-assembly, to name a few) has allowed
a veritable explosion of activity in this area. \cite{gen ref} It
is now well-known that carrier confinement into dimensions of a
few tens of nanometers provides strong blue shift of the
photoluminescence features from that in the original bulk
material, a clear consequence of quantum confinement in these
{\em quantum dots}. Currently, researchers are investigating a
variety of magnetocapacitance and optical properties of dots,
\cite{gen ref} including the role of inelastic light scattering
and phonon confinement, \cite{Jackson,Trallero} as well as Pauli
blocking and other few- and many-particle effects in these
systems. \cite{PawelNature}

In recent work, however, a new geometry of semiconductor quantum
{\em rings} has been introduced in experiments of
magnetocapacitance and infrared excitation for few-electrons.
\cite{LorkePRL00,GarciaAPL97} Although {\em metallic} rings have
been the subject of considerable attention for a number of years,
\cite{rings-rev} this geometry had not been achieved in
semiconductors for sizes such that the electrons would propagate
coherently (non-diffusively) throughout the ring. However, the
self-assembled quantum rings now achieved are so small (with
characteristic inner/outer radius of 20/100 nm and 2-3 nm in
height), that they allow the study of a non--simply-connected
geometry where carriers are coherent all throughout.  It is clear
that not only the single-particle states are interesting in this
geometry (specially their behavior under magnetic flux), but also
the role of interactions between particles (be it electrons or
holes). Lorke {\em et al}. have shown that multiple-electron
states in this geometry experience phase coherent effects in the
presence of magnetic fields, \cite{LorkePRL00} much as predicted
by theory. \cite{Tapash-rings}  The question of the observability
of similar coherent effects for electrons {\em and} holes around
the loop, in the presence of a magnetic field, is fascinating, and
some theoretical works have already begun to explore this regime.
\cite{Krive,Chaplik,Raikh,11'} Although beautiful experiments of
optical emission in charge-tunable quantum rings have been
recently presented, \cite{WarburtonNature,copycatters} they study
the role of multiply-charged exciton complexes with no applied
magnetic field.

The multiply-connected geometry of the semiconductor quantum
rings adds an interesting dimension to the strong Coulomb effects
in magnetic fields which have been explored in quantum confined
systems.  Excitons in magnetic fields have been investigated in
structures such as quantum wells, \cite{Maan,Cen} quantum wires,
\cite{Kohl} and quantum dots,
\cite{Halonen,Hawrylak,Song,Pedro,20'} as example of
multi-carrier systems. For the ring geometry, one question that
arises naturally is whether there is sensitivity of the exciton
properties to an applied flux. This `Aharanov-Bohm effect' (ABE)
for an exciton is an interesting concept, since one would
naturally associate the existence of the ABE with a net charge
(as the coupling constant to the vector potential), and the net
charge of this object is clearly zero. However, one could argue
that the composite nature of the excitons would perhaps allow for
a non-vanishing effect. In fact, for the case of particles
constrained to move along a one-dimensional ring, rigorous
derivations predict a non-zero ABE for the exciton, which will
show in its various energy states and the associated dipole
oscillator strength, for small enough rings. \cite{Chaplik,Raikh}

In this work, we present calculations of the excitation spectrum
and oscillator strength of excitons in rings pierced by magnetic
fields. We explore the role of different confinement potentials
and calculate binding energies, exciton sizes and their
dependence on magnetic fields, as well as oscillator strengths
which would be measurable in photoluminescence experiments, for
example. Similar to the case of quantum dots, we find strong
orbital effects from the magnetic field, which provides for
effectively stronger confinement and accompanying diamagnetic
level shifts, as well as splitting of some levels.  These changes
are found to be monotonic with field. Part of our motivation in
this study is to explore the question of how robust is the ABE
predicted in 1D rings, when one considers finite width and
confinement potentials.  The models we use are designed to mimic
the situation in real semiconductor quantum rings achieved to
date. Much to our chagrin, we find no trace of the predicted ABE
for realistic values of the rings and fields in the problem.
Although this negative result would suggest that it is difficult
that experiments would measure this effect, it is still open to
see to what extent highly sensitive experiments might be able to
yield a positive result.

The remainder of this paper is organized as follows. In Sec.\ II,
we present the model for the system and solution method. As a
first approximation, the quantum ring structure is modeled by a
parabolic confinement potential, in which the wavefunctions
expressed in terms of the center of mass and relative coordinates
are used as a basis set.  This confinement potential has been
experimentally confirmed by the recent experiments of Lorke {\em
et al}. \cite{LorkePRL00}.  In order to explore the role of
different potentials, we also model the ring system with a
hard-wall confinement, using wavefunctions expressed in terms of
Bessel functions for electron and hole as a basis set. In Sec.\
III, we discuss the main effects of the magnetic field effects on
the exciton characteristics, including the binding energy,
electron-hole separation, and the linear optical susceptibility.
Finally, we summarize our results in Sec.\ IV.  The Appendix
contains an outline of the derivation of the Coulomb matrix
element with these basis functions.  The analytical expressions
presented there greatly simplify our calculations.

 \section{Theoretical Model}

Our model is a two-dimensional exciton in a quantum ring and in a
static magnetic field, simulating recent experimental quantum
ring structures. The presence of magnetic fields oriented along
the $z$ axis, perpendicular to the plane of the ring, induces the
electron and hole to perform classical orbits along the
circumference, which of course yield quantization of the angular
momentum in that direction. \cite{Chaplik,Raikh} The ring
structures are well approximated by using parabolic potentials,
giving soft confinement barriers, appropriate to samples produced
by self-assembly. \cite{LorkePRL00}  For narrow rings (with steep
confinement), however, the parabolic confinement and associated
wavefunctions fail in a real system, as the increased confinement
may push the levels into the anharmonic part of the potential, and
even produce deconfinement of one carrier (typically the
electron), as found in some calculations in quantum dots.
\cite{Zunger}  We also consider the case of hard-wall confinement
and analyze the different results.

As the quantum rings and excitonic states under consideration are
much larger than the unit cell of the material, the effective-mass
approximation is a suitable approach, and is given by $ H = H_e +
H_h + H_{e-h} $, where the subscripts $e$ and $h$ represent
electron and hole, and the last term is the electron-hole
interaction. The expression for the Hamiltonian of the electron
in magnetic fields (in a parabolic-band approximation) is given by
\begin{equation}
 H_e = \frac{1}{2m_e}\left[{\bf p}_{e}+\frac{q_e}{c}{\bf
A}_{e}\right]^2 + V_e \, , \label{c4eq1}
\end{equation}
with a similar expression for the Hamiltonian of the hole. Here,
$V_e$ (or $V_h$) is the ring confinement potential for electron
(hole), and naturally $q_e=-|e|$ and $q_h=+|e|$. For parabolic
confinement potential across the width of the ring, we use
\begin{equation}
V_i = \frac{1}{2} m_i \omega _i^2 (r_{i}-r_{o})^2  \, ,
\label{c4-1eq2a}
\end{equation}
where the mean radius of the ring is $r_o$, and the characteristic
confinement energy is $\hbar \omega _i$, giving a characteristic
ring width $\approx 2 \sqrt {\hbar / m_i \omega_i}$ for each
particle.  Here, $i= e,h$ represents the different particles. We
choose the fully symmetric gauge vector potentials $ {\bf
A}_e=\frac{1}{2} {\bf B} \times ({\bf r}_e - {\bf r}_h)$ and ${\bf
A}_h=\frac{1}{2}{\bf B} \times ({\bf r}_h - {\bf r}_e)$, for
electron and hole, respectively, following Ref.\
[\onlinecite{Halonen,Pedro}]. The Coulomb interaction term between
carriers is given by $ H_{e-h} = - e^2 / \epsilon r_{e-h} $,
screened by the background dielectric constant $\epsilon$.

 \subsection{Parabolic confinement potential}

For the parabolic confinement potential is convenient to separate
the problem into center of mass and relative coordinates,
described as usual by ${\bf r} = {\bf r}_e - {\bf r}_h$, and
${\bf R} = (m{_e}{\bf r}_e + m_h{\bf r}_h)/M$, where the total
and reduced masses are $M = m{_e} + m{_h}$, and $\mu =
m{_e}m{_h}/M$. The total Hamiltonian can then be re-expressed as
$H = H_{CM} + H_{rel} + H_{mix} $,
with individual terms
\begin{eqnarray}
H_{CM} &=& \frac{1}{2M}P^2 +  \frac{1}{2} M \omega_o^2
(R-r_{o})^2 \, , \label{c4-1eq4} \\
H_{rel}^{o} &=& \frac{1}{2\mu}p^2 + \frac{1}{2} \mu \omega^2 r^2
\, , \label{c4-1eq5}
\end{eqnarray}
and $H_{rel} = H^o_{rel} + H^{\prime}_{rel} \,$, where
\begin{eqnarray}
H_{rel}^{\prime} &=& - \omega_c \gamma L_z - \frac{e^2}{\epsilon
r} \, , \label{c4-1eq6} \\
\mbox{{\rm and}} \nonumber \\
H_{mix} &=&  - \frac{e}{Mc} ({\bf B} \times {\bf r}) {\bf
{\cdot}} {\bf P}
\nonumber \\
&&- m_e \omega_o^2 r_o R \left(1+ \frac{2m_h}{MR^2} {\bf R}\cdot
{\bf r} + \frac{m_h^2}{M^2} \frac{r^2}{R^2} \right)^{1/2}
\nonumber \\
&&- m_h \omega_o^2 r_o R \left( 1- \frac{2m_e}{MR^2} {\bf R}\cdot
{\bf r} + \frac{m_e^2}{M^2} \frac{r^2}{R^2} \right)^{1/2} \, ,
\label{c4-1eq7}
\end{eqnarray}
where $\gamma =({m_h} - m_e) / M$ depends on the mass asymmetry
of the carriers, and we have set $\omega_e = \omega_h = \omega_o$.
We have also denoted the relative angular momentum in the
$z$-direction as $L_z = ({\bf r} \times {\bf p})_z$, and the
effective confinement frequency as $\omega^2 = \omega_o^2 +
\omega_c^2$, with $\omega_c = e B/(2\mu c)$, resulting from the
combined confinement of the potential and the magnetic field.

The main purpose in the change of variables above is to use the
solutions of $H_{CM}$ and $H_{rel}^{o}$ as a basis for the
solution of the  full Hamiltonian. The center of mass basis is
essentially a harmonic oscillator, with wavefunction
$\psi_{N,l_{CM}}$ centered about $r_o$,
 \begin{eqnarray}
 \psi_{N,l_{CM}} (R) &=& \alpha \sqrt{\frac{2 N!}{(N+|l_{CM}|)!}}
\frac{1}{\sqrt{2\pi}} e^{il_{CM}\theta} e^{-\alpha^2 (R-r_o)^2/2}
\nonumber \\
&\times& \left[ \alpha (R-r_o) \right]^{|l_{CM}|} L_N^{|l_{CM}|}
\left( \alpha^2(R-r_o)^2 \right) \, , \label{c4-1eq8}
 \end{eqnarray}
with $\alpha = \sqrt{M \omega _o / \hbar}$, and eigenvalues
$E_{\rm{CM}}^o = \left( 2N + 1 + |l_{CM}| \right) \hbar \omega_o$.
Here, $N$ and $l_{CM}$ are quantum numbers of the radial and
angular momentum part, respectively, for the center of mass
coordinates, and $L_{N}^{|l_{CM}|}$ is the associated Laguerre
polynomial. \cite{Hand}  Similarly, the eigenvalues and
eigenfunctions for the {\em non}-interacting relative Hamiltonian
are given by wavefunction $\phi_{n,l}$ and energy $E_{rel}^o$,
where
 \begin{eqnarray}
 \phi_{n,l} (r) &=& \beta \sqrt{\frac{2 n!}{(n+|l|)!}}
 \frac{1}{\sqrt{2\pi}} e^{il\varphi}
e^{-\beta^2 r^2/2} \left( \beta r \right)^{|l|} L_n^{|l|} \left(
\beta^2 r^2 \right) \, , \nonumber \\
E_{\rm{rel}}^o &=& \left( 2n + 1 + |l| \right) \hbar \omega \, ,
\label{c4-1eq11}
 \end{eqnarray}
with $\beta = \sqrt{\mu \omega / \hbar}$. Here, $n$ and $l$ are
quantum numbers of the radial and angular momentum part,
respectively, for the relative coordinates. With this harmonic
basis set, matrix elements for the Coulomb interaction, magnetic
field dependent and mixing terms can be calculated analytically.
\cite{Song,Pedro} These expressions are collected in the Appendix.

The total Hamiltonian given above is then diagonalized
numerically, leading to the eigenvalues and eigenfunctions. All
the physical properties of the exciton in the ring can in
principle be extracted from these eigenvalues and eigenfunctions.
Here, we present the binding energy, electron-hole separation,
and the linear optical susceptibility of the quantum ring. These
quantities are readily accessible in optical experiments of
photoluminescence (PL) and photoluminescence excitation (PLE).
Denoting the wavefunctions of the exciton as $|\Psi \rangle =
\sum_{Nl_{CM}nl} a_{Nl_{CM}nl} | N,l_{CM},n,l, \rangle$ with
coefficients $a_{Nl_{CM}nl}$ obtained from the diagonalization,
one can write for example an expression for the electron-hole
separation,
 \begin{eqnarray}
r_s^2 &=& \langle \Psi | r^2 | \Psi \rangle =
\delta_{N^{\prime},N} \,  \delta_{l_{CM}^{\prime},l_{CM}} \,
\delta_{l^{\prime},l} \,  \sum_{Nl_{CM}nl
N^{\prime}l_{CM}^{\prime}n^{\prime}l^{\prime}} a^{\ast}_{
N^{\prime}l_{CM}^{\prime}n^{\prime}l^{\prime}} a_{Nl_{CM}nl}
\nonumber \\
&\times&
\sqrt{\frac{n! \, \, n^{\prime}!} {(n+|l|)! \, (n^{\prime} + |l^{\prime}|)!}}
\sum_{k=0}^{n} \sum_{j=0}^{n^{\prime}} \frac{(n+|l|)! \,
(n^{\prime}+|l^{\prime}|)!}{(k+|l|)! \, (j+|l^{\prime}|)!}
\nonumber \\
&\times&
(-1)^{k+j} \frac{1}{k! \, (n-k)! \,  j! \,  (n^{\prime}-j)!}
\frac{1}{\beta ^2} \left( |l|+k+j+1 \right) ! \, .
 \label{c4-1eq14}
 \end{eqnarray}
Similarly, the linear optical susceptibility is given by $\chi (
\omega ) = \sum_{j} \left|\langle 0 | P | 1 \rangle_j \right|^2
\left( \hbar \omega -E_j -i \hbar \Gamma \right) ^{-1}$, where
$\langle 0 | P | 1 \rangle_j$ is the dipole matrix element
between one electron-hole pair $j$ state and the vacuum state.
These are proportional to the bulk interband matrix elements,
$p_{cv}$, and can be written in terms of the envelope function
as, \cite{Hermann}
\begin{eqnarray}
\left|\langle 0 | P | 1 \rangle\right|^2 &=& \left|p_{cv}\right|^2
\left|\int \Psi({\bf r}_e,{\bf r}_h={\bf r}_e) d{\bf r}_e
\right|^2
\nonumber \\
&=& \left| p_{cv} \right|^2 \left\{ \delta_{l,0} \sum_{n} a_n
\sqrt{\frac{\mu \omega}{\pi \hbar}} \right\}^2 \left\{
\delta_{l_{CM},0} \sqrt{2\pi} \sqrt{\frac{2 N!}{(N+|l_{CM}|)!}}
(-1)^N \frac{1}{{\alpha}} \right\}^2 \, .
\nonumber \\
\label{c4-1eq16}
\end{eqnarray}

\subsection{A hard-wall confinement potential}

Given that the two-dimensional free exciton size (effective Bohr
radius) in InAs is $a^{2D}_B \approx 16$ nm (6 nm for GaAs), the
quantum rings with widths larger than $2a_B^{2D}$ would tend to
yield highly symmetric (nearly circular) ground states of the
exciton, with the confinement potential being a small
perturbation. For narrower quantum rings, however, the symmetry
of the exciton in the ring would be strongly affected, and become
increasingly one-dimensional.  This would be favorable for the
appearance of the ABE, as predicted by theory.
\cite{Chaplik,Raikh} To allow for this different case, and so as
to test the possible bias of the numerical calculations in the
parabolic potential, we have also implemented solutions of the
problem in a hard-wall confinement potential basis.

In that situation, the basis set for the exciton problem is given
by products of the radial and angular parts of electron and hole,
$\Psi = \Psi_{e}(r_{e},\phi_{e}) \Psi_{h}(r_{h},\phi_{h})$, where
the individual wavefunctions are given by (in the absence of
magnetic field)
\begin{equation}
 \Psi(r_{i},\phi_{i}) = \psi_i(r_{i}) \frac{1}{\sqrt {2 \pi}} e^{il_{i}
 \phi_{i}} \, ,
\label{c4eq3}
\end{equation}
where $i=e,h $, and $l_i$ is an integer. The wavefunctions of the
radial part must satisfy the hard-wall boundary conditions and
vanish at both the inner ($a$) and outer radius ($a+2L$) of the
ring structure.  As such, they are given by linear combinations of
Bessel functions, $ \psi_i(r_i) = A J_{l_i}(k_ir_i) + B
N_{l_i}(k_ir_i) $, for $a \leq r_i \leq a+2L$. Here, $A$ and $B$
are normalized constants, and $J_{l_i}$ and $N_{l_i}$ are the
$l_i$th-order Bessel function of the first and second kind,
respectively, with $ k_i \, = \, \sqrt{2m_iE_i/ \hbar ^2}$. The
eigenvalue conditions are obtained from the secular equation
$J_{l_e}(k_ea) N_{l_e}(k_e(a+2L)) = N_{l_e}(k_ea)
J_{l_e}(k_e(a+2L)) $, with a similar expression for the hole
states. These expressions yield the basis for the electron-hole
pair problem without Coulomb interaction nor magnetic field, with
eigenvalues $E_o  =  \hbar^2 {k_e}^2/2m_e \, + \, {\hbar^2
{k_h}^2}/{2m_h}$. One can conveniently write the Coulomb
interaction potential matrix elements using this noninteracting
pair basis via Fourier transform integrals, as done in Ref.\
[\onlinecite{Tapash-rings}] (see Appendix).

Similarly, can obtain the matrix elements of the Hamiltonian
which depend on magnetic fields,
\begin{equation}
H_{B} = - \frac{e}{2m_ec} {\bf p}_e \cdot {\bf B} \times {\bf r}_e
+ \frac{e^2B^2}{8m_ec^2} {r_e}^2
+ \frac{e}{2m_hc} {\bf p}_h \cdot {\bf B} \times {\bf r}_h
+ \frac{e^2B^2}{8m_hc^2} {r_h}^2 \, ,
\label{c4eq20}
\end{equation}
by using straightforward finite domain integrals of the basis set
given above.  The energies and eigenfunctions for the exciton are
calculated by numerical diagonalization of the total Hamiltonian
{\em with} magnetic fields obtained from the summation of all
terms above. The wavefunctions are then represented as
$|\Psi\rangle = \sum_{n_en_hl_el_h} a_{n_en_hl_el_h} |
n_e,n_h,l_e,l_h \rangle $, where $a_{n_en_hl_el_h}$ are the
coefficients calculated from the diagonalization.  In turn, the
mean electron-hole separation $r_s$ and the linear optical
susceptibility can be calculated.

 \section{Results}

\begin{figure}
% \vskip -2cm
 \psfig{file=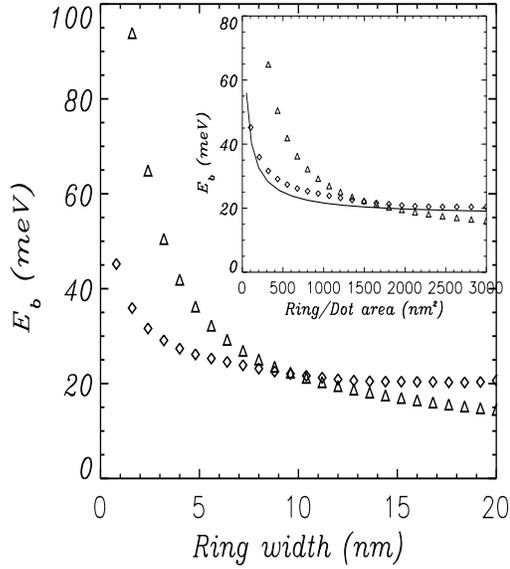,width=8cm,height=8cm,angle=0}
 \vspace{8ex}
 \caption{ Quantum ring heavy-hole exciton binding energies for
parabolic (triangles) and hard-wall (diamonds) confinement
potential, as function of the ring width. Inset: Same exciton
binding energies as function of the area for quantum rings
(triangles and diamonds traces), and quantum dot (solid line)
with the same area.} \label{fig1}
\end{figure}

We present here characteristic results of our calculations. As
mentioned before, they are scalable for different materials, in
terms of the Bohr radius of the exciton and its relation to the
size (specially the width) of the ring.  The parameters employed
here describe GaAs, yielding an effective 2D Bohr radius of 6 nm.
 Figure \ref{fig1} compares the exciton binding energies obtained
for parabolic confinement (triangles) with those for a hard-wall
confinement (diamonds), as function of the quantum ring width.
Notice that $E_b =E_{e-h}^0 - E_{grnd}^{exciton} $, where the
first term is only the confinement ground state of the electron
and hole, ignoring the Coulomb interaction.  For relatively narrow
rings, the exciton binding energy for the parabolic confinement
model is much larger than for the hard-wall confinement, as
expected. This difference is a reflection of the relative
strength of the kinetic energy to Coulomb attraction increasing
for the hard-wall case over the parabolic potential. In other
words, even though we use nominally the same width ($=
2\sqrt{\hbar/\mu \omega}$ for the parabolic potential and $2L$
for the hard wall), the parabolic potential solutions are
effectively less confined due to the finite amplitude `leaking'
out of the ring. It is also interesting to emphasize that for
smaller ring widths, the resulting wavefunctions are increasingly
elongated {\em along} the ring, and this is more the case for the
hard-wall confinement. In contrast, the exciton binding energy
appears larger for the hard-wall confinement, for widths larger
than $a_B^{2D}$, due to the known poor-convergence of the
parabolic basis used (e.g., see a detailed discussion of this
problem in Ref.\ \onlinecite{Song}). In the range of widths
shown, both approaches have converged numerically to within a few
percent everywhere, and at least an order of magnitude better for
the lower two-thirds of the range).

The inset in Fig.\ \ref{fig1} compares the ring results with those
of a quantum dot (with parabolic confinement \cite{Song}) with
equal confinement {\em area} (solid line). For wider ring systems
all energies are basically equal, as the confinement potential is
a weak perturbation to the Coulomb interaction between carriers,
as one would expect, be it ring or dot. On the other hand, the
exciton binding energies in the narrower rings are larger than in
the dot case with the same area, a reflection of the anisotropic
confinement in the ring: For the narrow rings, the circular
symmetry of the 2D free exciton (either free or in the dot) is
strongly affected, and the exciton elongates along the ring, as
described above.  We should mention that the {\em curvature} of
the ring has not much effect on the ground state or binding
energies for the dimensions considered here, similar to the
experimental values.

\begin{figure}
% \vskip -2cm
 \psfig{file=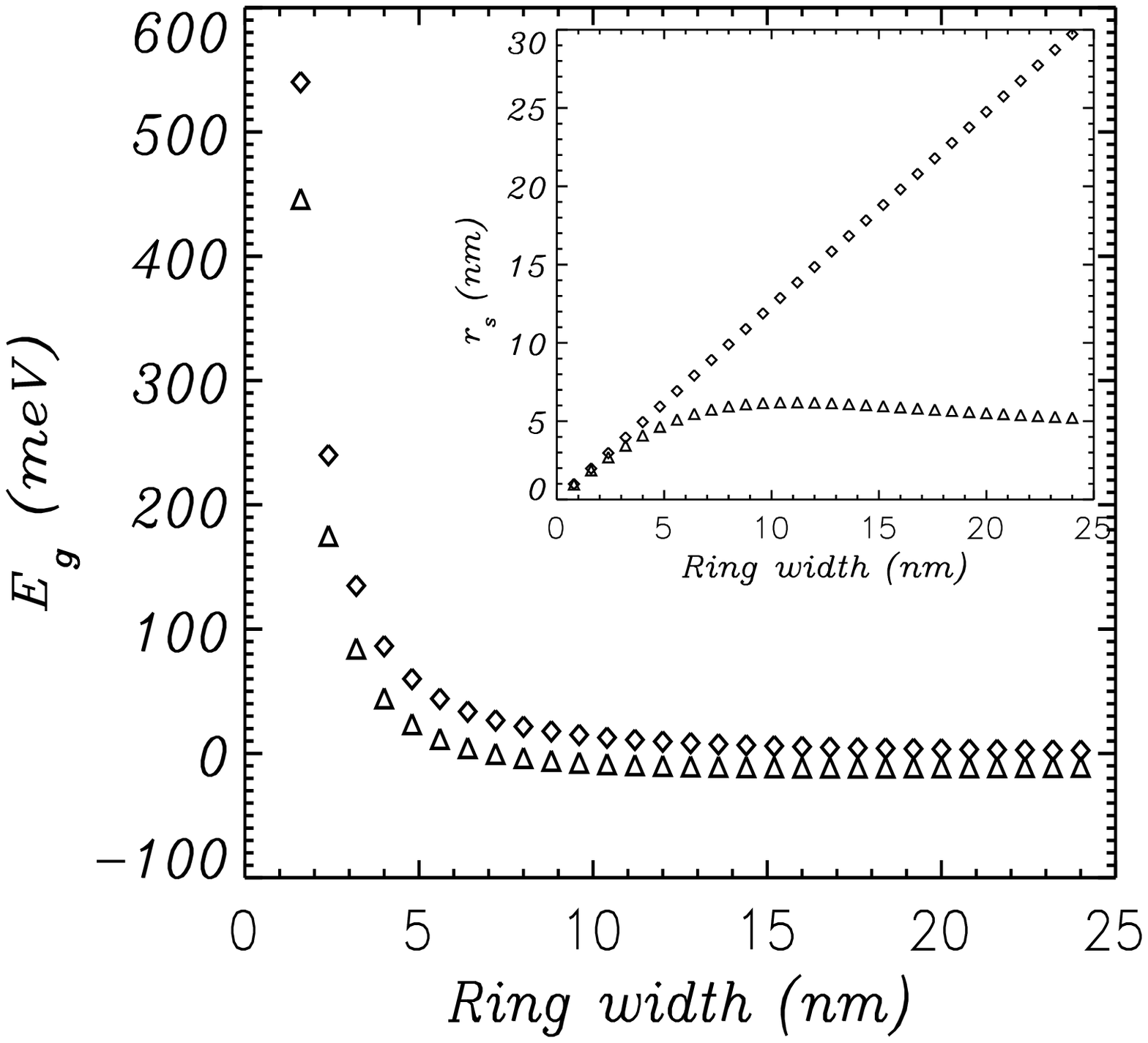,width=8cm,height=8cm,angle=0}
 \vspace{8ex}
 \caption{ Heavy-hole exciton ground state energy for parabolic
confinement, as a function of the ring width.  Inset:
Electron-hole separation vs.\ ring width.  Triangle and diamond
points are results for both with and without electron-hole
Coulomb interaction, respectively.} \label{fig3}
\end{figure}

To indicate the role of the Coulomb potential on the exciton
characteristics, Fig.\ \ref{fig3} shows the ground state energy
of the electron-hole pair in the parabolic-confinement ring with
(triangles) and without (diamond) Coulomb interaction. For a
smaller ring width, the Coulomb contribution clearly increases,
but not as fast as the confinement energy itself. The inset shows
the electron-hole separation vs.\ the ring width. For small
width, the confinement energy is clearly dominant in determining
the electron-hole separation, rather than the Coulomb interaction
term. For widths larger than $\approx 4$ nm, however, the
electron-hole separation depends mostly on the Coulomb
interaction term. Notice, however, that the rapid vanishing of
$r_s$ for thin rings is somewhat of a biasing artifact produced
by the basis functions we use in the parabolic ring. The
necessary truncation of the basis appears to favor a
circularly-symmetric exciton, clearly not the case in very thin
rings.

\begin{figure}
% \vskip -2cm
 \psfig{file=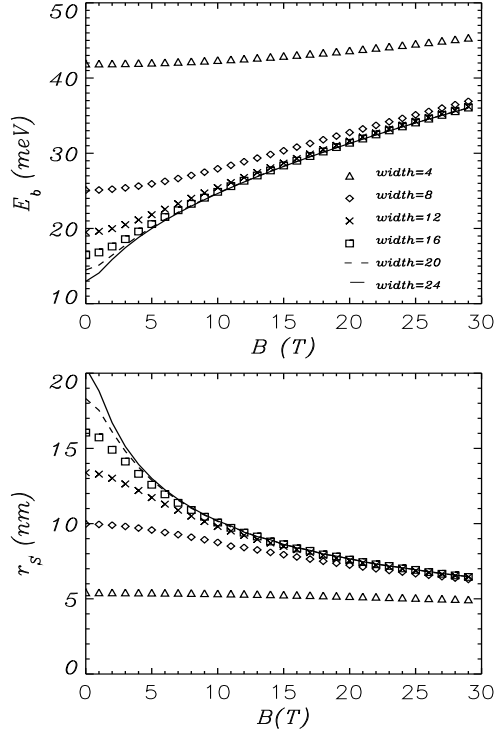,width=8cm,height=10cm,angle=0}
 \vspace{16ex}
 \caption{ Upper panel: Heavy-hole exciton binding energy as a function
of magnetic field for rings with $r_o=20$ nm.  Bottom panel:
Electron-hole separation as a function of magnetic field for same
rings.  Different symbols, as shown, indicate parabolic ring
width for heavy hole in nm.} \label{fig4}
\end{figure}

Figure \ref{fig4} shows the exciton binding energy and
electron-hole separation versus the external magnetic field, for
several ring widths, for a ring with middle radius $r_o = 20$ nm.
One can see that for larger values of the confinement energy
(i.e., smaller widths), the effect of the magnetic fields is
weaker, yielding the slowly changing curves at the top. However,
for the larger widths, the dependence of the exciton binding
energy on magnetic fields increases, resulting in the strong
enhancement of the binding energies and decreasing exciton sizes
with field. Notice that for larger values of the field, the
exciton binding energy changes little as function of the ring
width, showing that the confinement provided by the magnetic
field is dominant.  This is to be expected, given that the
magnetic length, $l_c = \sqrt{\hbar /\mu \omega_c}$ overtakes the
exciton radius at about 18 T.

For the radius of the ring in this figure, $r_o = 20$ nm, one
expects ABE oscillations with a periodicity given by multiples of
$B \pi r_0^2 / \phi_0$ (where $\phi_0 = hc/e = 4.14 \times
10^{-7}$ gauss$\cdot$cm$^2$ is the flux quantum), or a period
$\Delta B \approx 3.3$ T.\@  We find no appreciable evidence of
ABE oscillations in either the binding energy or the exciton
effective size, $r_s$. This result suggests that it is likely that
ABE exciton oscillations will not be seen in measurements of the
ground state properties of the exciton.

\begin{figure}
% \vskip -2cm
 \psfig{file=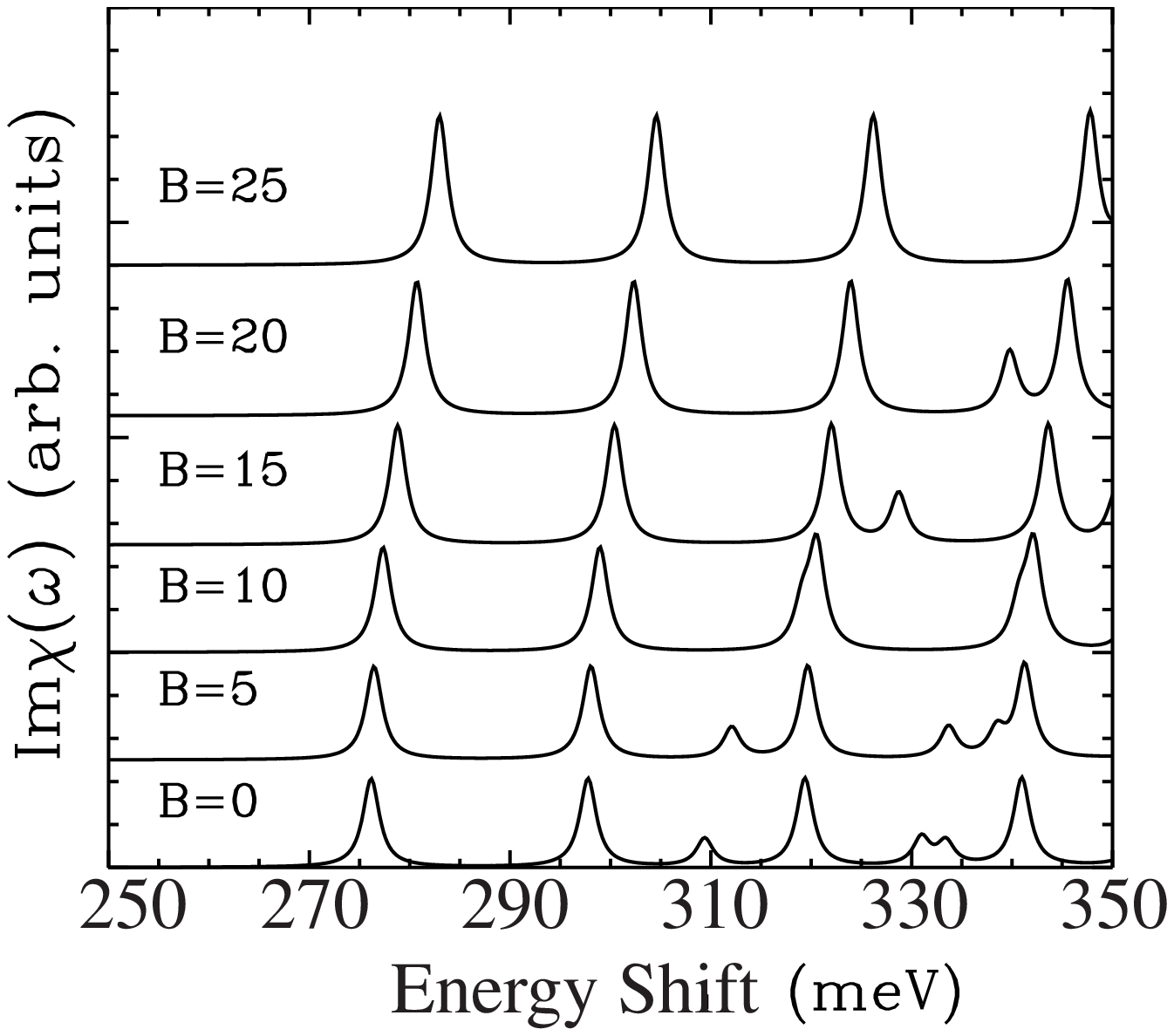,width=8cm,height=8cm,angle=0}
 \vspace{8ex}
 \caption{Linear optical susceptibility $\chi$ for parabolic quantum ring
with confinement energy $\hbar \omega _o = 10.8$ meV (exciton
$width=21$ nm) for magnetic fields ranging from $0$ to 25 T.\@
Radius of the ring is $24$ nm.  Energy blue shift includes both
in-plane and $z$-axis confinement ($z$-axis well width = 3nm).}
\label{fig5}
\end{figure}

Since there is a prediction that the ABE oscillations are to be
found much more easily in the case of excited states,
\cite{Raikh} we have also looked for them in the linear optical
susceptibility of a quantum ring.  Figure \ref{fig5} shows a
typical result for different values of the magnetic field. This
curve represents all the possible transitions of this excitonic
state which would be measurable via photoluminescence excitation
measurements (PLE; while the first peak gives the PL response).
The higher peaks, starting from the one at lowest energy,
correspond to electron-hole excitations involving the heavy-hole
exciton ground state and various center-of-mass replicas (i.e.,
increasing excitations of the center of mass degree of freedom,
without altering the ground state of the relative coordinate). On
the other hand, the smaller amplitude peaks (at shift $\approx
310$, 330 meV for $B=0$), correspond to {\em internal} excited
states of the exciton (its relative coordinate).  These peaks are
strongly upshifted with magnetic field, as the diamagnetic effect
for each charge carrier pushes all relative energies upwards as
well, and clearly these excited states shift even faster.  This
behavior is qualitatively similar to the excitons in a quantum
dot. \cite{Pedro}  We should also point out that in addition to
the overall upward shift due to the diamagnetic effect, the
$\chi$ traces show no discernible superimposed ABE oscillations
with magnetic field in any of the excited states.  It would
appear that the finite width of the system suppresses the ABE
predicted for the 1D ring.

We should also mention that higher excited states are likely to
exhibit the ABE effect, as per earlier work. \cite{Raikh}
However, the parabolic or hard-wall confinement models used in
this calculation lose validity, since non-parabolic corrections
to the effective mass hamiltonian, as well as finite confinement
potential effects would become more important.  Consequently, a
quantitative estimate of the anticipated ABE effects for
high-lying states is less reliable, and more subject to specific
values of parameters.  A weaker confinement might also enhance
ABE, although experiments in that regime would be harder to
identify and characterize uniquely.

 \section{Conclusions}

We have shown that magnetic field has strong effects on excitons
in a quantum ring, for both parabolic and hard-wall confinement
potentials. Using direct matrix diagonalization techniques, we
have shown that at least for rings currently realizable, the
excitons behave to a great extent as those in quantum dots of
similar dimensions: There are strong diamagnetic shifts and
restructuring of the overall excitation manifold, large shifts of
internal excitations, and reduction of the effective exciton
size.  On the other hand, the predicted ABE oscillations in the
various physical characteristics (including binding energy and
oscillator strength of transitions) of 1D excitons, are not found
in this more realistic calculation. Although we anticipated that
the predicted ABE effects would be much weaker (due to the finite
transverse size of the rings and ring radii larger than the
exciton size), we have not been able to detect any oscillation
`remnant', in any of the features we analyzed.

This negative result is due to either of two reasons, we surmise:
One, the result of exponentially small ABE amplitudes (given the
somewhat larger ring radii). \cite{Raikh} More likely, perhaps,
this is the result of destructive interference of relatively many
transverse eigenstates (mixed by the Coulomb interaction), each
with its own different phase and amplitude. Notice that this is
quite different for multiple-electron states in the ring, as
predicted by theory, and recently seen in
experiment.\cite{LorkePRL00} The difference in result from the
case of only electrons in the ring to that of an exciton,
indicates the predicted fragility of the effect, since in this
system the net charge (and then coupling to the magnetic vector
potential) is zero.  In fact, the delicate nature of the ABE
suggests that smaller and {\em narrower} rings are needed in
experiment, which may make for more one-dimensional--like
excitons. Following our discussion above, one could also expect
ABE oscillations to be more important for higher-excited states,
even if more challenging in experiments. Perhaps one could also
think of a technique that explores {\em differences}, and
therefore is able to couple only to a modulation of the hole
population, for example, as a sensitive way to access these
coherent ABE oscillations for low-lying states. \cite{NEWNOTE}

%*******************
   \acknowledgments

We would like to thank T. Shahbazyan, M. Cobb, A. Lorke, and C.
Trallero for helpful discussions.  This work has been supported in
part by the US Department of Energy grant no.\
DE--FG02--91ER45334.

\appendix
\section*{}

The matrix elements of $H_{rel}^{\prime}$, including the Coulomb
interaction term, can be calculated by using the parabolic ring
basis,
 \begin{eqnarray}
\langle N^{\prime} l_{CM}^{\prime} n^{\prime} l^{\prime} |
&H_{rel}^{\prime}&| N l_{CM} n l \rangle = - \delta_{N^{\prime},N}
\delta_{l_{CM}^{\prime},l_{CM}}
 \delta_{l^{\prime},l}  \left\{ \sqrt{\frac{n! \, \, n^{\prime}!}
{(n+|l|)! \, (n^{\prime} + |l^{\prime}|)!}} \right. \nonumber \\
&\times& \sum_{k=0}^n \sum_{j=0}^{n^\prime} \frac{(n+|l|)! \,
(n^{\prime}+|l^{\prime}|)!}{(k+|l|)! \, (j+|l^{\prime}|)!}
(-1)^{k+j} \frac{1}{k! \, (n-k)! \,  j! \,  (n^{\prime}-j)!} \nonumber \\
\times&& \left. \frac{e^2}{\epsilon} \beta \Gamma \left(
|l|+k+j+\frac{1}{2} \right) + \frac{eB\hbar l}{2\mu^{\prime}c}
\delta_{n^{\prime},n} \right\} \, . \label{c4-1eq12}
\end{eqnarray}

The matrix elements of the mixing terms between the center of
mass and relative coordinates, in the limit $r \ll R$, are of the
form
 \begin{eqnarray}
&&\langle N^{\prime} l_{CM}^{\prime} n^{\prime} l^{\prime} |
H_{mix}| N l_{CM} n l \rangle =
\nonumber \\
&\times&\sqrt{\frac{n! \, \, n^{\prime}!} {(n+|l|)! \,
(n^{\prime} + |l^{\prime}|)!}} \, 2 \alpha^2 \sqrt{\frac{N! \,
N^{\prime}!}{(N+|l_{CM}|)! (N^{\prime} + |l_{CM}^{\prime} |)!}}
\nonumber \\
&\times&\sum_{k=0}^n \sum_{j=0}^{n^\prime} \, \frac{(n+|l|)! \,
(n^{\prime}+|l^{\prime}|)!}{(k+|l|)! \, (j+|l^{\prime}|)!} \,
(-1)^{k+j} \, \frac{1}{k! \, (n-k)! \,  j! \,  (n^{\prime}-j)!}
\nonumber \\
&\times& \sum_{K=0}^N \sum_{J=0}^{N^\prime} \frac{(N+|l_{CM}|)! \,
(N^{\prime}+|l_{CM}^{\prime}|)!} {(K+|l_{CM}|)! \,
(J+|l_{CM}^{\prime}|)!} (-1)^{K+J} \frac{1}{K! \, (N-K)! \,  J! \,
(N^{\prime}-J)!}
\nonumber \\
&\times& \left[ \frac{e B \hbar}{Mc\beta}\Gamma \left(
\frac{2j+2k+|l| +|l^{\prime}|+3}{2} \right) \right. \left\{
\left[ - \frac{1}{4\alpha} \Gamma \left(
\frac{2J+2K+|l_{CM}|+|l_{CM}^{\prime}|+3}{2} \right)
\right. \right. \nonumber \\
&+& \frac{1}{4\alpha} \left( 2K+|l_{CM}| \right) \Gamma \left(
\frac{2J+2K+|l_{CM}|+|l_{CM}^{\prime}|+1}{2} \right)
\nonumber \\
&-&\frac{r_o}{4} \Gamma \left(
\frac{2J+2K+|l_{CM}|+|l_{CM}^{\prime}|+2}{2} \right) +
\frac{r_o}{4} \left( 2K+|l_{CM}| \right)
\nonumber \\
&\times&\left. \Gamma \left(
\frac{2J+2K+|l_{CM}|+|l_{CM}^{\prime}|}{2} \right)
 \right]
\left[ \delta_{l,l^{\prime}+1} \, \,  \delta_{l_{CM},l_{CM}
^{\prime}-1} - \delta_{l,l^{\prime}-1} \, \, \delta_{l_{CM},l_{CM}
^{\prime}+1} \right] \nonumber \\
&-& \left. \left[ \delta_{l,l^{\prime}+1} \,  \,
\delta_{l_{CM},l_{CM}^{\prime}-1} + \delta_{l,l^{\prime}-1} \, \,
\delta_{l_{CM},l_{CM}^{\prime}+1} \right]
\frac{l_{CM}}{4\alpha}\Gamma \left( \frac{2J+2K+|l_{CM}|
+|l_{CM}^{\prime}|+1}{2} \right)
 \right\}
\nonumber \\
&-& \frac{\mu \omega _o^2 r_o}{4\alpha \beta^2}
\delta_{l,l^{\prime}} \delta_{l_{CM},l_{CM}^{\prime}} \left(
|l|+k+j \right) ! \left\{ \Gamma \left( |l_{CM}|+K+J+\frac{1}{2}
\right)
\right. \nonumber \\
&-& \left. \left. e^{-\alpha ^2r_o^2} \sum_{jj=0}^{\infty} \frac{
\left( |l_{CM}| +K+J-\frac{1}{2} \right) !}{\left(
|l_{CM}|+K+J+\frac{1}{2}+jj \right) !} \left( \alpha ^2 r_o^2
\right)^{|l_{CM}|+K+J+\frac{1}{2}+jj} \right\} \right] \, .
\nonumber \\
\label{c4-1eq13}
\end{eqnarray}

Coulomb interaction in hard wall confinement,
\cite{Halonen,Pedro} is given by

\begin{equation}
 H_{e-h}({\bf q}) =  - \frac{e^2}{\epsilon}
 {(2\pi)^2} \int \frac{1}{{\bf r}_{e-h}} e^{-i{\bf q} \cdot
{\bf r}_{e-h}} d{\bf r}_{e-h}
 = - 2\pi \frac{e^2}{\epsilon} \frac{1}{q}  \, ,
\label{c4eq10}
\end{equation}
and
\begin{equation}
\Psi_e({\bf r}_e) = \frac{1}{(2\pi)^2} \int \phi_e({\bf q})
e^{-i{\bf q} \cdot {\bf r}_e} d{\bf q}  \, . \label{c4eq11}
\end{equation}

The Coulomb interaction matrix elements,
\begin{equation}
\langle n_e^{\prime}l_e^{\prime}n_h^{\prime}l_h^{\prime} |\,
H_{e-h}\, | n_el_en_hl_h \rangle = \int \Psi_e^{\prime}({\bf r}_e)
\Psi_h^{\prime}({\bf r}_h) H_{e-h}({\bf r}_{e-h}) \Psi_e({\bf
r}_e) \Psi_h({\bf r}_h) d{\bf r}_e d{\bf r}_h \, , \label{c4eq12}
\end{equation}
are then rewritten as
\begin{eqnarray}
\langle n_e^{\prime}l_e^{\prime}n_h^{\prime}l_h^{\prime} |\,
H_{e-h}\, | n_el_en_hl_h \rangle&=&\frac{1}{(2\pi)^6} \int
\phi_e^{\prime}({\bf q}_e)
\phi_h^{\prime}({\bf q}_h) H_{e-h}({\bf q}) \nonumber \\
&\times&\phi_e({\bf q}_e-{\bf q}) \phi_h({\bf q}_h+{\bf q}) d{\bf
q}_e d{\bf q}_h d{\bf q} \, , \label{c4eq13}
\end{eqnarray}
where the Fourier transform integrals used for electron and hole
wavefunctions, respectively, are
\begin{eqnarray}
\varphi_e({\bf q})&\equiv&\int \phi_e^{\prime}({\bf q}_e)
\phi_e({\bf q}_e-{\bf q}) d{\bf q}_e \nonumber \\ &=& (2\pi)
\int_{a}^{a+2L} \psi_{e}^{\prime}(r_e) \psi_{e}(r_e) r_e dr_e
\int_{0}^{2\pi} e^{im\phi_e} e^{-iqr_ecos(\phi_q-\phi_e)}
d\theta_e \nonumber \\  &=& (2\pi)^2 (i)^{-m} e^{im\phi_q}
\int_{a}^{a+2L} \psi_{e}^{\prime}(r_e) \psi_{e}(r_e) J_{m}(r_eq)
r_e dr_e \, , \label{c4eq14}
\end{eqnarray}
where $m=l_e-l_{e}^{\prime}$, and the definition of the Bessel
function
\begin{equation}
J_{m}(qr) \, = \, \frac{1}{2\pi} \int_{-\pi}^{\pi}
e^{-im\phi+izsin\phi} d\phi
\end{equation}
has been used.  A similar expression for the hole wavefunctions
gives the interaction matrix elements
\begin{eqnarray}
\langle n_e^{\prime}l_e^{\prime}n_h^{\prime}l_h^{\prime}
|&H_{e-h}&| n_el_en_hl_h \rangle = -
\delta_{l_e+l_h,l_{e}^{\prime}+l_{h}^{\prime}}
\frac{e^2}{\epsilon} \int_{a}^{a+2L} \psi_{e}^{\prime}(r_e)
\psi_e(r_e) r_e dr_e
\nonumber \\
&\times& \int_{a}^{a+2L} \psi_{h}^{\prime}(r_h) \psi_h(r_h) r_h
dr_h \int_{0}^{\infty} J_{|m|}(r_eq) J_{|m|}(r_hq) dq \, .
\label{c4eq16}
\end{eqnarray}
In order to evaluate the integrals in Eq.\ (\ref{c4eq16}), we use
\begin{eqnarray}
\int_{0}^{\infty} J_{|m|}(r_eq)J_{|m|}(r_hq) dq
 =&&\frac{r_{<}^{|m|}}
{r_{>}^{|m|+1}}
\frac{\Gamma(|m|+\frac{1}{2})}{\Gamma(|m|+1)\Gamma( \frac{1}{2})}
\nonumber \\
&\times& F\left(|m|+\frac{1}{2},\frac{1}{2};|m|+1;\left(
\frac{r_<}{r_>} \right)^2 \right) \, , \label{c4eq17}
\end{eqnarray}
where $r_>$ ($r_<$) is the larger (smaller) of $r_e$ and $r_h$,
and $F$ is a hypergeometric function. \cite{Integral-Hand}
Inserting (\ref{c4eq17}) into (\ref{c4eq16}), the interaction
matrix elements can be written as
\begin{eqnarray}
\langle n_e^{\prime}l_e^{\prime}n_h^{\prime}l_h^{\prime}
|&H_{e-h}&| n_el_en_hl_h \rangle = -
\delta_{l_e+l_h,l_{e}^{\prime}+l_{h}^{\prime}}
\frac{e^2}{\epsilon}
\frac{\Gamma(|m|+\frac{1}{2})}{\Gamma(|m|+1)\Gamma(\frac{1}{2})}
\nonumber \\
&\times&\left\{ \int_{a}^{a+2L} \psi_{e}^{\prime}(r_e) \psi_e(r_e)
\frac{1}{r_{e}^{|m|}} dr_e \right. \int_{a}^{r_e}
\psi_{h}^{\prime}(r_h) \psi_h(r_h) r_{h}^{|m|+1}
\nonumber \\
&   &\times \,
F\left(|m|+\frac{1}{2},\frac{1}{2};|m|+1;\left(\frac{r_h}{r_e}
\right)^2\right) dr_h
\nonumber \\
&+&\int_{a}^{a+2L} \psi_{h}^{\prime}(r_h) \psi_h(r_h)
\frac{1}{r_{h}^{|m|}} dr_h \int_{a}^{r_h} \psi_{e}^{\prime}(r_e)
\psi_e(r_e) r_{e}^{|m|+1}
\nonumber \\
&   &\times \, \left.
F\left(|m|+\frac{1}{2},\frac{1}{2};|m|+1;\left( \frac{r_e}{r_h}
\right)^2\right) dr_e \right\} \, . \label{c4eq18}
\end{eqnarray}
This greatly simplified expression for the interaction matrix
elements is easily evaluated numerically. The total Hamiltonian
matrix is diagonalized numerically, providing all the eigenvalues
and eigenfunctions. To improve numerical convergence, we use the
transformation of hypergeometric functions given by
\begin{eqnarray}
F\left(|m|+\frac{1}{2},\frac{1}{2};|m|+1;z\right) =
\frac{\Gamma(|m|+1)}{\Gamma(|m|+\frac{1}{2})\Gamma(\frac{1}{2})}
\sum_{n=0}^{\infty} \frac{(|m|+\frac{1}{2})_n (\frac{1}{2})_n}
{{n!}^2}
\nonumber \\
\times \, \left\{2 {\bf \psi}(n+1) - {\bf
\psi}(|m|+n+\frac{1}{2}) - {\bf \psi} (n+\frac{1}{2}) - ln(1-z)
\right\} (1-z)^n \, , \label{c4eq19}
\end{eqnarray}
where ${\bf \psi}$ is the digamma function. \cite{Special-Hand}

The matrix elements dependent on magnetic fields are given by
\begin{eqnarray}
&&\langle n_{e}^{\prime}l_{e}^{\prime}n_{h}^{\prime}l_{h}^{\prime}
|\, H_{B}\,  | n_el_en_hl_h \rangle = \delta_{l_{e}^{\prime},l_e}
\, \delta_{l_{h}^{\prime},l_h} \left\{ \left(\frac{l_h}{m_h} -
\frac{l_e}{m_e}\right) \frac{\hbar eB}{2c} \right. \,
\nonumber \\
&& \, \, \, \, \, \, \,   \times\int_{a}^{a+2L}
\psi_{e}^{\prime}(r_e) \psi_{e}(r_e) r_e dr_e \int_{a}^{a+2L}
\psi_{h}^{\prime}(r_h) \psi_{h}(r_h) r_h dr_h
\nonumber \\
&+&\, \, \frac{e^2B^2}{8m_ec^2} \int_{a}^{a+2L}
\psi_{h}^{\prime}(r_h) \psi_{h}(r_h) r_h dr_h \int_{a}^{a+2L}
\psi_{e}^{\prime}(r_e) \psi_{e}(r_e) {r_e}^3 dr_e
\nonumber \\
&+&\, \, \left. \frac{e^2B^2}{8m_hc^2} \int_{a}^{a+2L}
\psi_{e}^{\prime}(r_e) \psi_{e}(r_e) r_e dr_e \int_{a}^{a+2L}
\psi_{h}^{\prime}(r_h) \psi_{h}(r_h) {r_h}^3 dr_h \right\} \, .
\label{c4eq21}
\end{eqnarray}

The size of the exciton is here given by
\begin{eqnarray} {r_s}^2
&=& \langle \Psi | r^2 | \Psi \rangle = \sum_{n_en_hl_el_h
n_{e}^{\prime}n_{h}^{\prime}l_{e}^{\prime}l_{h}^{\prime}}
a_{n_{e}^{\prime}n_{h}^{\prime}l_{e}^{\prime}l_{h}^{\prime}}^{\prime}
a_{n_en_hl_el_h}
\nonumber \\
&\times& \left\{ \delta_{l_{e}^{\prime},l_e}
\delta_{l_{h}^{\prime},l_h} \left( \int_{a}^{a+2L}
\psi_{h}^{\prime}(r_h) \psi_{h}(r_h) r_h dr_h \right. \right.
\int_{a}^{a+2L}\psi_{e}^{\prime}(r_e) \psi_{e}(r_e) {r_e}^3 dr_e
\nonumber \\
&+& \left. \int_{a}^{a+2L} \psi_{e}^{\prime}(r_e) \psi_{e}(r_e)
r_e dr_e \int_{a}^{a+2L} \psi_{h}^{\prime}(r_h) \psi_{h}(r_h)
{r_h}^3 dr_h \right)
\nonumber \\
&-& \delta_{l_{e}^{\prime}+l_{h}^{\prime},l_e+l_h}
\delta_{l_{e}^{\prime}+1,l_e} \int_{a}^{a+2L}
\psi_{e}^{\prime}(r_e) \psi_{e}(r_e) {r_e}^2 dr_e
\nonumber \\
& &\times \left. \int_{a}^{a+2L} \psi_{h}^{\prime}(r_h)
\psi_{h}(r_h) {r_h}^2 dr_h \right\} \, . \label{c4eq22}
\end{eqnarray}

%\end{multicols}


\begin{references}
\bibitem{gen ref} See, for example, the proceedings of
International Conf. on Semiconductor Quantum Dots, QD2000,
Munich, Germany, to be published in Physica Status Solidi.

\bibitem{Jackson} S. Lee, J.C. Kim, H. Rho, C.S. Kim, L.M. Smith,
H.E. Jackson, J.K. Furdyna, M. Dobrowolska, Phys. Rev. B {\bf 61},
R2405 (2000).

\bibitem{Trallero} E. Menendez-Propin, C. Trallero-Giner, and S.E.
Ulloa, \prb {\bf 60}, 16 747 (1999).

\bibitem{PawelNature} M. Bayer, O. Stern, P. Hawrylak, S. Fafard,
and A. Forschel, Nature {\bf 405}, 923 (2000).

\bibitem{LorkePRL00}  A. Lorke, R.J. Luyken, A.O. Govorov, J.P.
Kotthaus, J.M. Garcia, P.M. Petroff, \prl {\bf 84}, 2223 (2000).

\bibitem{GarciaAPL97} J.M. Garcia, G. Medeiros-Ribeiro,
K. Schmidt, T. Ngo, J. L. Feng, A. Lorke, J. Kotthaus, and P. M.
Petroff, \apl {\bf 71}, 2014 (1997).

\bibitem{rings-rev} See, for example, M. B\"uttiker, Y. Imry,
and R. Landauer, Phys. Lett. {\bf 96A}, 365 (1983); L. P. Levy
{\em et al}., Phys. Rev. Lett. {\bf 64}, 2074 (1990); V.
Chandrasekhar {\em et al}., ibid {\bf 67}, 3578 (1991); D.
Mailly, ibid {\bf 70}, 2020 (1993).

\bibitem{Tapash-rings} T. Chakraborty and P. Pietil\..{a}inen,
Phys. Rev. B {\bf 50}, 8460 (1994); {\bf 52}, 1932 (1995).

\bibitem{Krive} I. Krive and A.A. Krokhin, Phys. Lett. A {\bf
186}, 162 (1994).

\bibitem{Chaplik} A. Chaplik, JETP Lett. {\bf 62}, 900 (1995).

\bibitem{Raikh} R.A. R\"omer and M.E. Raikh, The Aharanov-Bohm
effect for an exciton, condmat/9906314.

\bibitem{11'} A. Gracia-Cristobal and C. Trallero-Giner,
unpublished.

\bibitem{WarburtonNature} R.J. Warburton, C. Sch\"aflein, D. Haft,
F. Bickel, A. Lorke, K. Karrai, J.M. Garcia, W. Schoenfeld, and
P.M. Petroff, Nature {\bf 405}, 926 (2000).

\bibitem{copycatters} F. Findis, A. Zrenner, G. B\"ohm, and G.
Abstreiter, Solid State Commun. {\bf 114}, 227 (2000).

\bibitem{Maan} J. C. Maan, G. Belle, A. Fasolino, M. Altarelli, and
K. Ploog, Phys. Rev. B {\bf 30}, 2253 (1984).

\bibitem{Cen} J. Cen and K. K. Bajaj, Phys. Rev. B {\bf 46}, 15280
(1992).

\bibitem{Kohl} M. Kohl, D. Heitmann, P. Grambow, and K. Ploog, Phys.
Rev. Lett. {\bf 63}, 2124 (1989).

\bibitem{Halonen} V. Halonen, T. Chakraborty, and P. Pietil\"{a}inen, Phys.
Rev. B {\bf 45}, 5980 (1992).

\bibitem{Hawrylak} L. Jacak, P. Hawrylak, and A. Wojs, Quantum
Dots (Springer, Berlin, 1998), and references therein.

\bibitem{Song} J. Song and S.E. Ulloa, Phys. Rev. B {\bf 52},
9015 (1995).

\bibitem{Pedro} P. Pereyra and S.E. Ulloa, \prb {\bf 61}, 2128
(2000).

\bibitem{20'} K. L. Janssens, F. M. Peeters, and V. A.
Schweigert, cond-mat/0002405.

\bibitem{Zunger} See, for example, L.W. Wang, A.J. Williamson,
A. Zunger, H. Jiang, and J. Singh, Appl. Phys. Lett. {\bf 76},
339 (2000), and references therein.

\bibitem{Hand} {\em Handbook of Mathematical Functions}, Page 374,
M. Abramowitz and I. A. Stegun (Dover, New York, 1972).

\bibitem{Hermann} C. Hermann and C. Weisbuch, Phys. Rev. B {\bf 15},
823 (1977).

\bibitem{Integral-Hand} {\em Table of Integrals, Series and Products}
Page 693, I. S. Gradshteyn and I. M. Ryzhik (Academic Press, Inc.
San Diego, 1994).

\bibitem{Special-Hand} {\em Special Functions \& Their Applications}
Page 6, N. N. Lebedev (Dover New York, 1972).

\bibitem{NEWNOTE} {\em Note added after review}: Since submission, we
have become aware of a preprint where calculations for rings
suggest a positive result for the appearance of ABE oscillations
[H. Hu, J.-L. Zhu, D.-J. Li, and J.-J. Xiong, cond-mat/0010310].
Our results here are not in contradiction with those, we believe,
since their somewhat unrealistically narrow rings enhance the ABE
oscillations, as we discuss. Those results in fact validate our
suspicion that narrower rings, if possible, are needed for these
effects to be seen.

\end{references}
\end{document}